\documentclass[%
twocolumn,
superscriptaddress,
notitlepage,
showpacs,
amsmath,amssymb,
aps,
prb,
floatfix
]{revtex4-2}

\usepackage{amsmath}
\usepackage{amssymb}

\usepackage{graphicx}
\usepackage{xcolor}
\usepackage{hyperref}

\usepackage{braket}
\usepackage[thinc]{esdiff}

\begin{document}

	\title{Influence of Charge Density Waves on the Hall coefficient in NiTi}
	\author{Adrian Braun}
	\affiliation{
		Physics Department, Bielefeld University, Postfach 100131, 33501 Bielefeld, Germany
	}
	\author{Henrik Dick}
	\affiliation{
		Physics Department, Bielefeld University, Postfach 100131, 33501 Bielefeld, Germany
	}
	\author{Timon Sieweke}
	\affiliation{
		Institute for Energy and Materials Processes – Applied Quantum Materials, University Duisburg-Essen, 47057 Duisburg, Germany
	}
	\author{Alexander Kunzmann}
	\affiliation{
		Institute for Energy and Materials Processes – Applied Quantum Materials, University Duisburg-Essen, 47057 Duisburg, Germany
	}
	\author{Klara L\"unser}
	\affiliation{
		Institute for Energy and Materials Processes – Applied Quantum Materials, University Duisburg-Essen, 47057 Duisburg, Germany
	}
 	\author{Gabi Schierning}
	\affiliation{
		Institute for Energy and Materials Processes – Applied Quantum Materials, University Duisburg-Essen, 47057 Duisburg, Germany
	}
        \author{Thomas Dahm}\email{thomas.dahm@uni-bielefeld.de}
	\affiliation{
		Physics Department, Bielefeld University, Postfach 100131, 33501 Bielefeld, Germany
	}
	
	\date{\today}
	
	\begin{abstract}
          We present a mean-field charge density wave theory for NiTi using density functional theory bandstructure
          as a starting point. We calculate the Hall coefficient as a function of temperature and compare with
          recent experimental results. We analyze the contributions to
          the Hall coefficient from different parts of the Fermi surface and find that the Hall coefficient is dominated by certain
          ``hot spots''. The analysis shows that these hot spots are mostly dominated by Ni $d$-orbitals. We demonstrate that the Hall coefficient
          is not well reproduced by Boltzmann transport theory within the constant relaxation time approximation without charge density waves.
          We consider both uniaxial and biaxial charge density waves and show that biaxial charge density waves can account well
          for the Hall coefficient, while uniaxial cannot. We also investigate the temperature dependence of the resistivity and the specific heat. 
	\end{abstract}
	\maketitle
	\section{Introduction}
	\label{sec:intro}
        The shape memory effect in NiTi is due to a highly reversible martensitic structural phase transition \cite{OtsukaRen2005,Gruenebohm2023}. A general understanding of
        the structural change has been achieved by molecular dynamics (MD) simulations either based on effective interatomic
        potentials \cite{Mutter2010,KoMEAM2015} or by ab-initio molecular dynamics (AIMD) \cite{WuLawson2022}.
        However, a precise theoretical understanding of the transition temperature and the electronic properties of NiTi is still lacking.
        Molecular dynamics is based on the Born-Oppenheimer approximation \cite{Born1927} which assumes that the electrons can instantanously follow
        the motions of the nuclei. It thus neglects nonadiabatic effects like the retarded electron-phonon interaction, which is crucial
        to understand instabilities in the electronic system like charge density waves (CDWs), for example.
        There have been speculations about the possible formation of CDWs in NiTi already during the 1980s \cite{PapaKammSSC1982,SalamonPRB1985}.
        Arguments in favor of this conjecture have been superlattice reflections in X-ray, neutron and electron diffraction \cite{SalamonPRB1985,Todai2011}, the appearence
        of a gap like feature in the optical conductivity and a suppression of the magnetic
        susceptibility in the martensitic phase \cite{Shabalovskaya1985}, Fermi surface nesting features seen in band structure calculations \cite{PapaKammSSC1982,ZhaoPRB1989, Katsnelson2010},
        and phonon softening at a wave vector unrelated to the martensite phase \cite{Tietze1984,ZhaoHarmon1993}.
        But also some doubts were raised \cite{ShapiroPRB1984}.
        Direct observation of a CDW by scanning tunneling microscopy (STM) or the CDW gap by angular resolved photoemission spectroscopy (ARPES)
        has been difficult due to the insufficient surface quality of this material.
        Also, to our knowledge a quantitative theoretical study on the appearance of CDWs in NiTi has never been presented, which would allow
        a more direct comparison between experiment and theory.
        
        Recent experimental results on electronic transport properties of this material found a strong decrease
        of the charge carrier density and a strong increase of carrier mobility
        in the martensite phase from Hall coefficient measurements. Also, a surprisingly large contribution
        of the electronic part of the entropy change at the phase transition was found \cite{Kunzmann2022,KunzmannDiss},
        hinting towards a strong involvement of the electronic subsystem in the phase transition. In the present work we investigate
        a mean-field theory of the formation of different types of CDWs in NiTi based on ab-initio bandstructure.
        We calculate the Hall coefficient within this model and compare with the experimental data
        in an effort to gauge the possibility of a CDW forming in this material.

  	\begin{figure}
		\includegraphics[width=1.0\linewidth]{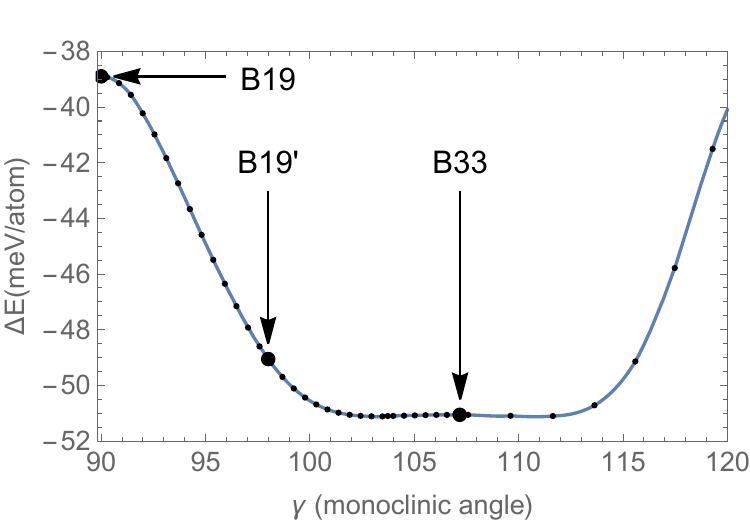}
		\caption{Ground state energy relative to the B2 phase as a function of monoclinic angle.}
		\label{fig:energyvsgamma}
	\end{figure}

        \begin{figure*}[ht]
		\includegraphics[width=0.9\linewidth]{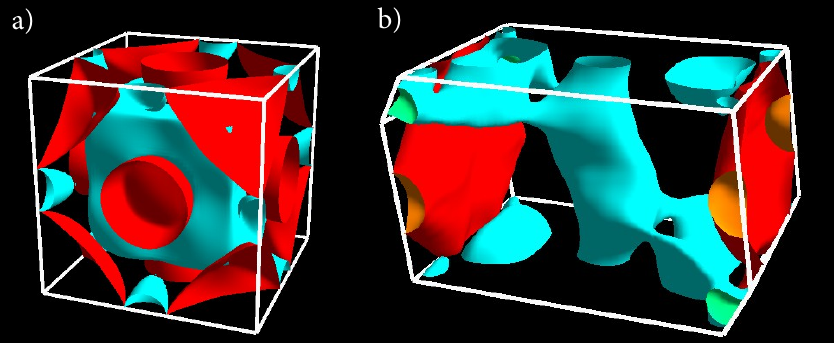}
		\caption{Fermi surfaces of (a) the B2 and (b) the B19' phase.}
		\label{fig:fermisurfaces}
	\end{figure*}

        In section \ref{sec:NiTiDFT} we will review the current understanding of the structural phase transition in NiTi based on ab-initio molecular dynamics and introduce
        the bandstructures we are using. In section \ref{sec:BTP2} we will present Boltzmann transport calculations based on these bandstructures and compare them with
        measurements of the Hall coefficient and resistivity. Our CDW mean-field theory is introduced in section \ref{sec:CDW} and compared with experimental results.
        
	\section{Current understanding of the structural phase transition in NiTi}
	\label{sec:NiTiDFT}
        NiTi undergoes a martensitic phase transformation from a high temperature simple cubic austenite B2 phase into a low temperature monoclinic martensite B19' phase
        with a monoclinic angle of 98° \cite{Buhrer1983}. There has been a long-standing discussion about the structure of the zero-temperature ground state
        of NiTi, as DFT calculations tended to yield an orthorhombic phase, the B33 phase, to be the most stable one, which is inconsistent with the
        observed shape memory effect, however \cite{Huang2003,Haskins2016,Kumar2020,WuLawson2022}.
        This puzzle has been resolved recently by Wu and Lawson, who could show that both the finite temperature entropic effects
        together with the zero-point energy of the phonons conspire to make the B19' phase more stable than the B33 phase \cite{WuLawson2022}.
        The martensistic transition temperature still was found to be 80~K above the experimental one, however.
        
        In Fig.~\ref{fig:energyvsgamma} we present our own DFT calculation of the ground state energy as a function of monoclinic angle.
        We have used the QuantumEspresso package \cite{QE_2009,QE_2017} within the generalized gradient approximation using the more recent revised
        Perdew-Burke-Ernzerhof approximation for solids (PBEsol) \cite{PBEsolPRL2008}. For the ultrasoft pseudopotentials we used an energy cutoff of 100~Ry.
        Mazari-Vanderbilt cold smearing of 0.02~Ry was used \cite{MVcoldsmearing} together with a $8 \times 8 \times8$ Monkhorst-Pack $k$-point grid. The lattice constants and the atomic positions within the unit cell
        were fully relaxed under the constraint of a fixed monoclinic angle. The ground state energies shown in Fig. ~\ref{fig:energyvsgamma} are in very good
        agreement with previous calculations \cite{Huang2003,WuLawson2022}. In contrast to previous work, we find a minimum of the energy at an monoclinic
        angle of 103.5°, however. We attribute this to our use of the PBEsol pseudopotential. We note that the energy curve becomes extremely flat in the range
        between $\gamma=99$° and $\gamma=114$°, where the variation is below 1~meV/atom. The minimum at 103.5° is found to be just 0.06~meV/atom below the energy of the B33 phase.
        This result demonstrates that the position of the energy minimum is highly sensitive to small changes in the pseudopotential due to the flatness of the
        energy curve. We did not include the phonon zero point energy here, which is expected to shift the minimum further down to 101° \cite{WuLawson2022}.

        As the experimentally observed monoclinic angle of the B19' phase is 98°\cite{Buhrer1983}, in the following we will restrict our discussion of electronic transport properties
        to a comparison of the cubic B2 phase and the B19' phase. We will not aim to understand the transition region between these two phases, which is a hysteretic
	first order transition, but instead try to understand the behavior in the two limiting homogeneous phases.

\begin{figure}
		\includegraphics[width=1.0\linewidth]{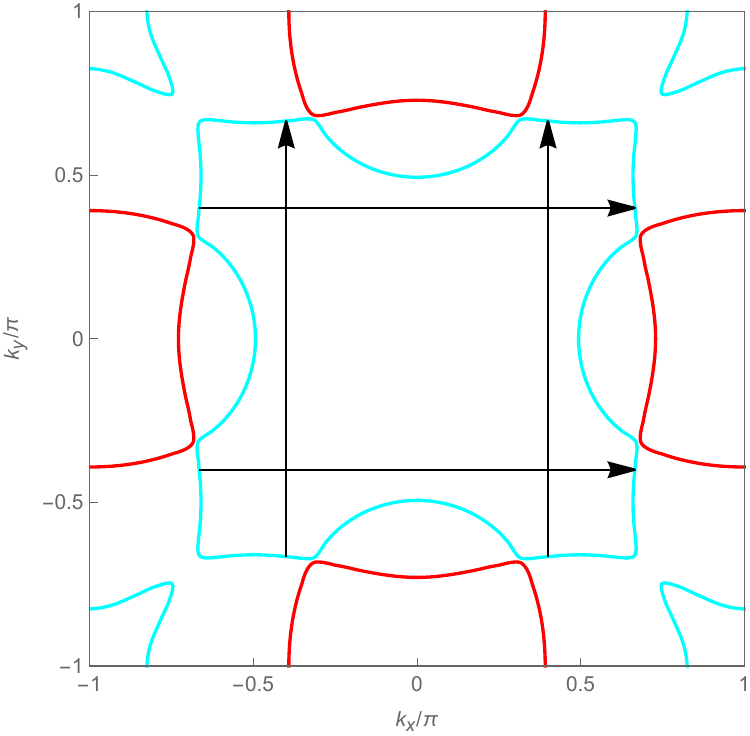}
		\caption{Cut of the Fermi surface in the $k_z=0$ plane. The arrows show the vectors $Q_1=\frac{4}{3} (\pi,0,0)$ and $Q_2=\frac{4}{3} (0,\pi,0)$.}
		\label{fig:cutFS}
        \end{figure}

        The Fermi surfaces in B2 and B19' phases are shown in Fig.~\ref{fig:fermisurfaces}, which are in good agreement with previous results
	\cite{PapaKammSSC1982,ZhaoPRB1989,Hatcher2009,Katsnelson2010}. In the B2 phase there are two bands crossing the Fermi surface, denoted
	here in blue and red color. Centered around the $\Gamma$ point one can see a large cube-like structure in blue with red ``bowls'' near the six surfaces of the cube. In the corners of the Brillouin zone
        red sail-like structures are visible, which touch the corner of the cube, and near the $M$-point blue pockets can be seen, which touch the sails. In Fig.~\ref{fig:cutFS} a cut through the 
        the Fermi surface at $k_z=0$ is shown, which demonstrates that the bowls intersect the cube in an avoided band crossing. Thus, there are several points in the Brillouin zone where the two bands
        cross right at the Fermi surface. In the B19' phase (Fig.~\ref{fig:fermisurfaces}b) there are four bands crossing the Fermi surface. The highest and lowest band of these four bands are only barely crossing the Fermi level
        leading to an electron and a hole pocket at the Brillouin zone boundary shown here in green and orange color. The two main structures are a tubular network (blue) around the $\Gamma$-point
        and a pancake-like structure (red) on the left and right end of the Brillouin zone. While the B2 Fermi surface possesses the full cubic symmetry, in the B19' phase there are
	only an inversion symmetry and one mirror symmetry.

	\begin{figure}
		\includegraphics[width=1.0\linewidth]{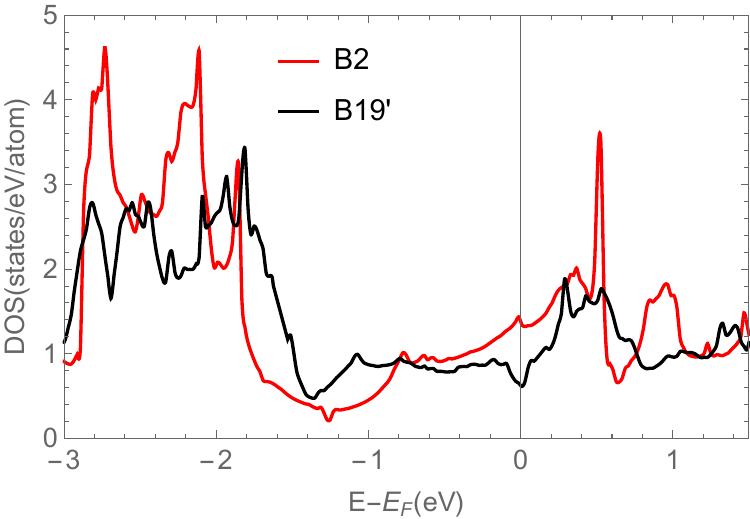}
		\caption{Density of states in the B2 phase (red) and the B19' phase (black).}
		\label{fig:DOS1}
	\end{figure}

        The density of states per atom (DOS) in the two phases is shown in Fig.\ref{fig:DOS1}. Here it becomes apparent that the density of states at the Fermi level is
        about a factor of two smaller in the B19' phase. The electronic transport properties and thermal properties are determined by the band structure in the close
        vicinity of the Fermi level. In particular the Fermi surface, the Fermi velocity and the band curvature at the Fermi level play a dominant role. Fermi velocity 
	and band curvature 
        vary significantly along the Fermi surface and cannot be considered constant. We calculated the band structure on a $16 \times 16 \times 16$ Monkhorst-Pack $k$-point grid.
        However, this grid is not fine enough when it comes to the calculation of transport properties, which needs to compute integrals over the Fermi surface.
        The usual technique is to interpolate the bands and do the Fermi surface integrals on a much finer grid to obtain sufficient accuracy. The popular code
        {\it BoltzTrap2} \cite{BTP2,BTP1}, for example, uses a smoothed Fourier interpolation to do the job. Another popular code, {\it Wannier90} \cite{Wannier90}
        constructs maximally localized Wannier functions from the Bloch states. These Wannier functions can then be used to create a tight-binding model of the
        bandstructure that provides a smooth Wannier interpolation
        of the bands \cite{MarzariReview}. Both methods have their limitations, however. While the smoothed Fourier interpolation is very fast, it sometimes
        creates oscillatory behavior near band crossings. This makes the method less reliable at low temperatures. This is a particular problem in the present case, as
        we have several band crossings right at the Fermi level, as pointed out above. The construction of the maximally localized
        Wannier functions takes much more time and for metals a disentanglement of the bands is necessary, which introduces a systematic error into the bands.
        Two of us have recently proposed another method to obtain accurate tight-binding models from DFT bandstructures, which is inspired by machine-learning techniques \cite{Dick2025}.
        This method is fast, its accuracy can be systematically controlled, and it does not suffer from oscillations. We have chosen to use this method here, as we could not
        get sufficient accuracy at low temperatures from the other two methods. A comparison of the DFT bandstructure with this tight binding representation can be seen
        in Fig. 3e in Ref.~\onlinecite{Dick2025}.

        In the following section~\ref{sec:BTP2} we present our Boltzmann transport calculations for the two phases B2 and B19' and compare them with our experimental results
        on NiTi samples.

        \section{Boltzmann transport calculations and comparison with experiment}
        \label{sec:BTP2}

\begin{figure}
		\includegraphics[width=1.0\linewidth]{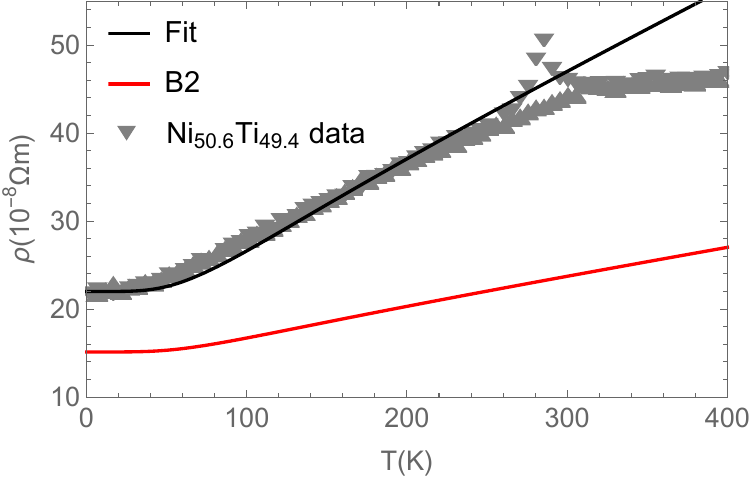}
		\caption{Temperature dependence of the resistivity compared with the Bloch-Gr\"uneisen fit in the B19' phase (black line).
                The red line is a hypothetical resistivity in the B2 phase (see text). The upward pointing triangles are our measured data
		  for increasing temperature and downward pointing triangles for decreasing temperature. }
		\label{fig:rhovsT_BGfit}
        \end{figure}

\begin{figure}
		\includegraphics[width=1.0\linewidth]{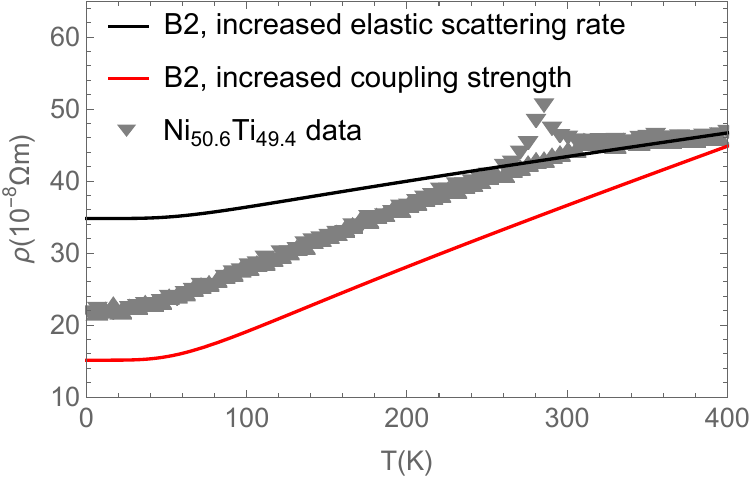}
		\caption{Temperature dependence of the resistivity in the B2 phase, if an increased elastic scattering rate is assumed (black line)
                  or an inceased electron-phonon coupling strength is assumed (red line).}
		\label{fig:rhovsT_scenarios}
        \end{figure}

\begin{figure}
		\includegraphics[width=1.0\linewidth]{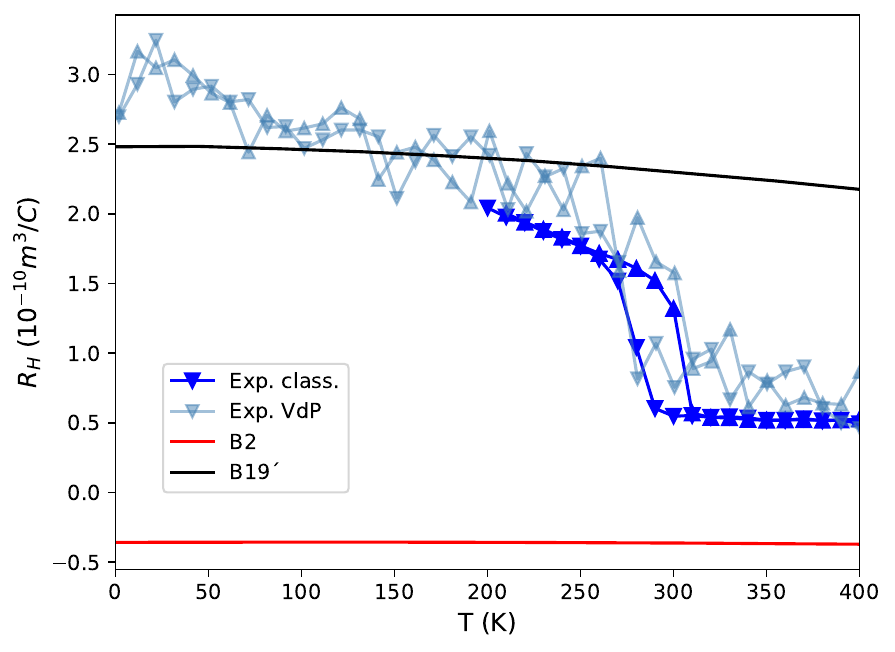}
		\caption{Temperature dependence of the Hall coefficient. The blue triangles show experimental data measured on a Ni$_{50.6}$Ti$_{49.4}$ sample by
                  two different methods: a classical Hall measurement (dark blue) and a van der Pauw measurement (light blue). The upward pointing triangles are
		  for increasing temperature and downward pointing triangles for decreasing temperature. The theoretical calculations
                within Boltzmann transport theory and constant relaxation time approximation are shown for the B19' phase (black line) and the B2 phase (red line).}
		\label{fig:RHall_normal}
        \end{figure}

        The sample used for Hall and resistance measurements shown in Figs. \ref{fig:rhovsT_BGfit}, \ref{fig:rhovsT_scenarios}, and \ref{fig:RHall_normal}
        was grown using the micro-pulling-down (µPD) method \cite{Fukuda2007}. In this process, NiTi is molten in a graphite crucible and
        slowly pulled through a hole at the bottom of the crucible using a NiTi seed wire. Prior to the micro–pulling-down process, the pure elements Ni and Ti were arc molten
        in an optimized procedure to achieve polycrystalline Ni$_{50}$Ti$_{50}$ as described in Ref.~\onlinecite{Frenzel2007} as feed material. The µPD grown sample (v = 60 mm/h at approx. 1366 °C) was cut
        perpendicular to the growth direction and polished to a thickness of 0.1 mm. A different piece from the same sample was used to estimate the Ni$_{50.6}$Ti$_{49.4}$ ratio
        via differential scanning calorimetry using the relation to the transition temperature \cite{Frenzel2010}. A detailed description of the home-made micro–pulling-down set-up, the
        parameters of the growing process, as well as chemical microstructural properties of comparable NiTi crystals can be found in Ref.~\onlinecite{Sieweke2026}.
        Resistance and Hall measurements were performed using a Physical Property Measurement System, 9T Dynacool (Quantum Design). All measurements were carried out using
        alternating current and a lock-in amplifier. Classical Hall measurements were conducted at magnetic fields of ±8 T, ±4 T and 0 T with a current of 5 mA.
        Van der Pauw (VdP) Hall and resistance measurements were performed with a switch box using a current of 3 mA at magnetic fields of ±4 T, ±2 T and 0 T \cite{Luenser2025}.

        Using our tight-binding representation of the bandstructure we have calculated the isotropic Hall coefficient $R_H$ and the Drude weight $D=\sigma/\tau$ within
        Boltzmann transport theory. For a given bandstructure $\epsilon_{\mathbf{k},n}$ the group velocity is given by
        \begin{equation}
          v_{\mathbf{k}n,\alpha} = \frac{1}{\hbar} \frac{\partial}{\partial k_{\alpha}} \epsilon_{\mathbf{k}n}
          \label{eq:group_volocity}
        \end{equation}
        where $n$ is the band index. The inverse mass tensor is obtained from
        \begin{equation}
          M_{\mathbf{k}n,\beta \mu}^{-1} = \frac{1}{\hbar^2} \frac{\partial^2}{\partial k_{\beta} k_{\mu}} \epsilon_{\mathbf{k}n} \; .
          \label{eq:inverse_mass_tensor}
        \end{equation}
        Note that these two quantities can be calculated analytically from the tight-binding model and evaluated at any desired $\mathbf{k}$-point within the
        first Brillouin zone. The rank-2 conductivity tensor can be calculated from
         \begin{equation}
          \sigma_{\alpha \beta} = \sum_{\mathbf{k}, \sigma, n} e^2 \tau_{\mathbf{k}n} v_{\mathbf{k}n,\alpha} v_{\mathbf{k}n,\beta} \left( - \frac{\partial}{\partial \epsilon} f(\epsilon_{\mathbf{k}n}) \right)
          \label{eq:rank2_conductivity}
         \end{equation}
         and the rank-3 conductivity tensor \cite{BTP1,Hurdbook,Schulz1992} from
         \begin{equation}
           \sigma_{\alpha \beta \gamma} = \sum_{\mathbf{k}, \sigma, n} e^3 \tau_{\mathbf{k}n}^2 \varepsilon_{\gamma \mu \nu} v_{\mathbf{k}n,\alpha} v_{\mathbf{k}n,\nu} M_{\mathbf{k}n,\beta \mu}^{-1}
           \left( - \frac{\partial}{\partial \epsilon} f(\epsilon_{\mathbf{k}n}) \right)
          \label{eq:rank3_conductivity}
         \end{equation}
         Here, $\tau_{\mathbf{k}n}$ is the scattering time, $f(\epsilon)=\frac{1}{1+e^{\left( \epsilon-\mu \right)/k_BT}}$ the Fermi function, and the summations run over the first Brillouin zone.
         The rank-3 Hall tensor is then obtained from
          \begin{equation}
           R_{\alpha \beta \gamma} = (\sigma^{-1})_{\mu \beta} \sigma_{\mu \nu \gamma} (\sigma^{-1})_{\alpha \nu}
          \label{eq:rank3_Hall_tensor}
         \end{equation}
          Within the constant relaxation time approximation one assumes that $\tau_{n, \mathbf{k}}=\tau=$~const. In this case $\sigma_{\alpha \beta} \propto \tau$ and
          $\sigma_{\alpha \beta \gamma} \propto \tau^2$ and as a result $R_{\alpha \beta \gamma}$ becomes independent of the scattering time.
          The isotropic Hall coefficient $R_H$ and the Drude weight $D$, valid for polycrystalline samples, are obtained from
        \begin{equation}
          R_H = \frac{1}{6} R_{ijk} \varepsilon_{ijk}
          \quad \text{ and } \quad 
          D = \frac{tr\{\sigma_{\alpha \beta} \} }{3 \tau}
          \label{eq:Hall_Drude}
        \end{equation}
        More details on the numerical evaluation of the above expressions can be found in the supplementary information.

        The Drude weight only has a very weak temperature dependence. However, it possesses significantly different values in the B2 and the B19' phases.
        In the B2 phase at $T=275 K$ in the vicinity of the transition temperature we find $D=7.32 \cdot 10^{20}/\Omega m s$, while in the B19' phase we have $D=2.51 \cdot 10^{20}/\Omega m s$,
        which is almost a factor of 3 smaller. In order to compare these values with the experimental resistivity $\rho=1/(D \tau)$, the temperature dependence of the scattering time $\tau$
        is needed. Here we are using the standard Bloch-Gr\"uneisen theory for electron-phonon scattering in metals and a constant elastic scattering rate due to impurities and
        imperfections in the sample:
        \begin{equation}
          \frac{1}{\tau} =  \frac{1}{\tau_{el}} + \frac{1}{\tau_{ph}}
          \label{eq:scattering_rate}
        \end{equation}
        where the phonon scattering rate is given by
        \begin{equation}
          \frac{1}{\tau_{ph}} = \frac{2 \pi k_B T}{\hbar} \lambda_{tr} B(T)
          \label{eq:phonon_scattering_rate}
        \end{equation}
        Here, $\lambda_{tr}$ is the dimensionless electron-phonon transport coupling strength and $B(T)$ a dimensionless temperature dependent function, which is normalized such that
        $B(T) \rightarrow 1$ for $T \rightarrow \infty$ \cite{Mazin1984}.
        Within Bloch-Gr\"uneisen theory
        \begin{equation}
          B(T) = 4 \left( \frac{T}{\Theta_D} \right)^4 J_5 \left( \frac{\Theta_D}{T} \right) 
          \label{eq:Bloch_Grueneisen}
        \end{equation}
        where $\Theta_D$ is the Debye temperature and
        \begin{equation}
          J_5(x) = \int_0^x dy \; y^5 \frac{e^y}{\left( e^y - 1 \right)^2}
          \label{eq:J5}
        \end{equation}
        \cite{Czycholl2023}.
        
        We obtained the Debye temperature $\Theta_D=370$~K by a fit to our specific heat data, which is shown in the supplementary information.
        This value is in good agreement with previous experimental and theoretical results \cite{Niitsu2015,Liu2023}.
        We use the two unknowns $\frac{1}{\tau_{el}}$ and $\lambda_{tr}$ in Eq.~(\ref{eq:scattering_rate}) and Eq.~(\ref{eq:phonon_scattering_rate}) 
        to obtain a two parameter fit to our experimental data in the B19' phase. The result is shown in Fig.~\ref{fig:rhovsT_BGfit} as the black line.
        The martensitic phase transition occurs in this sample between 270~K and 300~K and is visible as a peak feature in the resistivity.
        The fit can reasonably reproduce the data in the martensite phase below 260~K and is obtained for $\frac{\hbar}{\tau_{el}}=37$~meV and $\lambda_{tr}=0.28$. This suggests a sizable amount of
        elastic scattering and an intermediate strength of the electron-phonon coupling. Using the calculated Drude weight we can find a hypothetical resistivity curve for the B2 phase.
        For that we assume that the elastic scattering rate is proportional to the density of states at the Fermi energy \cite{Czycholl2023} and we assume that the electron-phonon coupling strength
        remains the same. The resulting curve is shown as the red line in Fig.~\ref{fig:rhovsT_BGfit}. Apparently these assumptions do not fit well to the resistivity in the austenite
        phase above 300~K. It can mean that the elastic scattering rate is significantly larger in the B2 phase or the electron-phonon coupling strength is significantly larger, or both. As an illustration
        in Fig.~\ref{fig:rhovsT_scenarios} we show the temperature dependence of the resistivity in the B2 phase, if we either increase the elastic scattering rate by a factor of 2.3 (black line) or
        if we increase the electron-phonon coupling strength by a factor of 2.5 to $\lambda_{tr}=0.7$ (red line). In both cases the resistivity in the austenite phase is reached.
        It seems unlikely, however, that the elastic scattering rate increases in the austenite phase.
        In the martensite phase there exists a large amount of twin boundaries, which are expected to create more elastic scattering than in the austenite phase.
        On the other hand, an increase of the electron-phonon coupling strength to $\lambda_{tr}=0.7$ in the B2 phase may be possible, in principle. A clarification of this
        issue has to await detailed calculations of elastic scattering rates and electron-phonon interaction in the B2 phase. We wish to point out that such calculations are difficult,
        because the B2 phase has unstable phonon modes at zero temperature which are entropically stabilized at higher temperatures \cite{Katsnelson2010,WuLawson2022}.
        Thus, standard techniques to calculate the electron-phonon interaction like density functional perturbation theory (DFPT) \cite{Baroni1987,Savrasov1994,Heid2013} or the
        frozen phonon technique \cite{Dacorogna1985,Risueno2023} need to be extended to account for the entropic effects, which has not been developed so far.
        
        \begin{figure*}
		\includegraphics[width=0.9\linewidth]{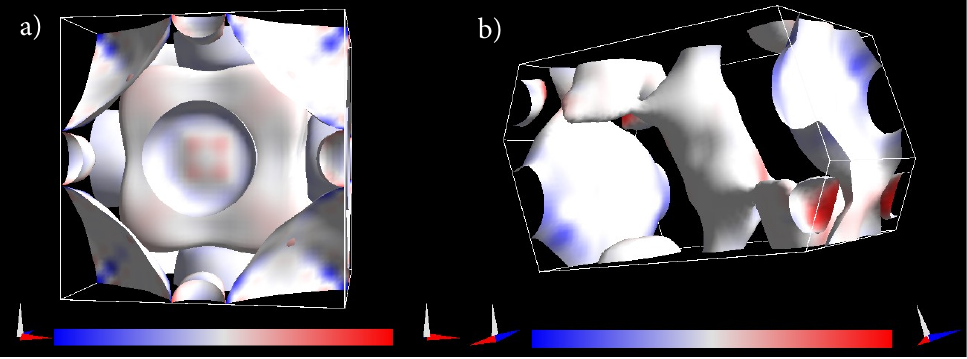}
		\caption{Momentum resolved contributions to the Hall coefficient shown on the Fermi surface of (a) the B2 and (b) the B19' phase.
			Blue color denotes negative, white color zero, and red color positive contributions.}
		\label{fig:Hall_contributions}
	\end{figure*}

        \begin{figure*}
		\includegraphics[width=0.9\linewidth]{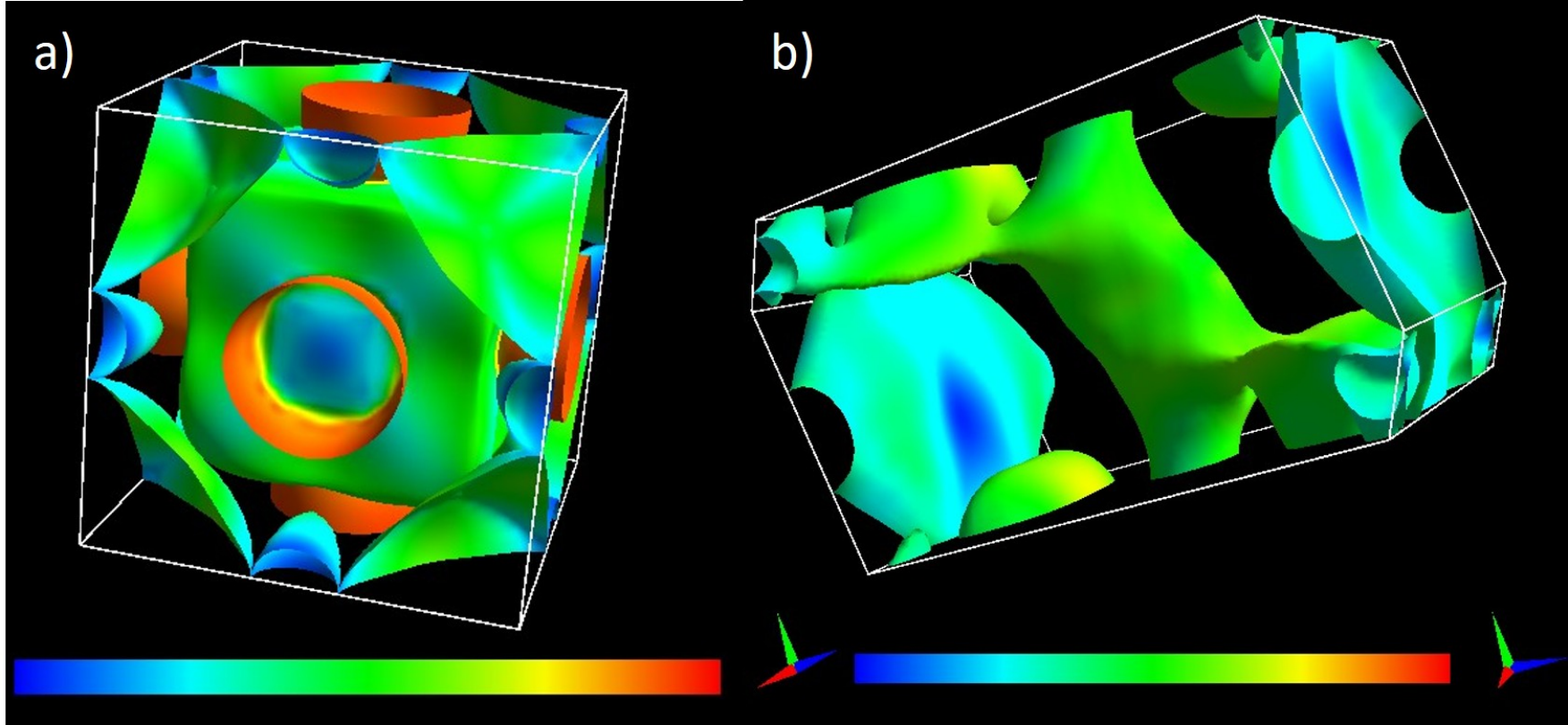}
		\caption{Orbital projection of the Fermi surface states onto Ti orbitals in the (a) B2 and (b) B19' phase.
                Here, dark blue corresponds to 10\%, green to 50\% and red to 90\% Ti contribution.}
		\label{fig:Ti_projections}
	\end{figure*}

        We next discuss our calculations and measurements of the Hall coefficient $R_H$. Within the constant relaxation time approximation
        the scattering time is cancelled out in the expression for the Hall coefficient. Thus, the Hall coefficient is not affected by the scattering
        issues just discussed for the resistivity, at least to lowest order in the momentum fluctuations of the scattering time. In Fig.~\ref{fig:RHall_normal}
        we present the experimental results for the same sample as in Fig.~\ref{fig:rhovsT_BGfit} using two different methods (classical Hall measurement
        and van der Pauw measurement). The black line shows the theoretical calculation in the B19' phase and the red line the one in the B2 phase.
        First of all one notes, that the experimental result below 270~K shows a clear increase of the Hall coefficient with decreasing temperature.
        The theoretical calculations in contrast show only a weak temperature dependence, which is a usual behavior in metallic systems as the
        temperature dependence of $R_H$ is usually governed by the high energy scale of the electronic system. In the B19' phase there is a rough
        agreement of the values of $R_H$ in experiment and theory. In the B2 phase, however, the theoretical calculation gives a wrong sign of the
        Hall coefficient. In order to elucidate the reason for this sign change, we have calculated the momentum resolved contributions to the
        Hall coefficient. In Fig.~\ref{fig:Hall_contributions} we show them as a color scale on top of the Fermi surfaces in the B2 and B19' phases.
        Here, blue color denotes a negative contribution, white color zero contribution, and red color a positive contribution.
        Interestingly, most of the Fermi surfaces appear white, giving only a weak contribution to the Hall coefficient. In the B2 phase
        (Fig.~\ref{fig:Hall_contributions}a) strong negative contributions can be seen near the corners of the sail-like parts of the Fermi surface,
        while some weak positive contributions are visible on some other parts of the Fermi surface. In the B19' phase (Fig.~\ref{fig:Hall_contributions}b)
        there are strong positive contributions coming from the pocket-like features of the Fermi surface and some negative contributions seen on the
        pancake-like structure. In both cases certain spots on the Fermi surface dominate the Hall coefficient. In the B2 phase the negative contributions
        are prevaling, while in the B19' phase the positive contributions dominate.

        \begin{figure}
		\includegraphics[width=1.0\linewidth]{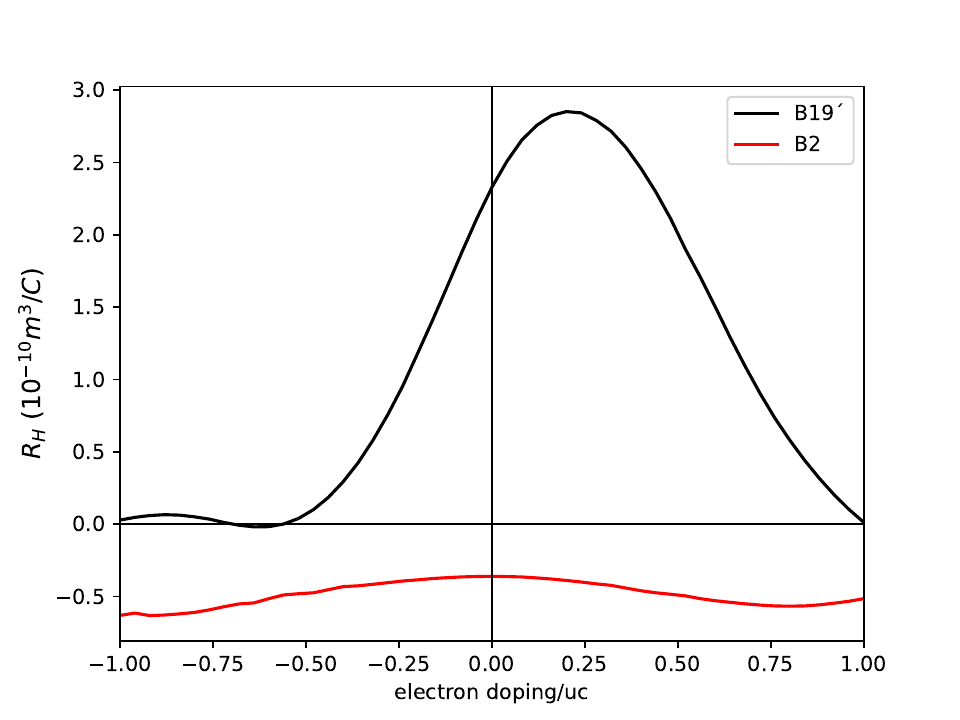}
		\caption{Hall coefficient at 275~K as a function of electron doping in both B2 (red) and B19' (black) phases.}
		\label{fig:RH_vs_doping}
        \end{figure}

        The states at the Fermi surface mostly consist of Ni and Ti $d$-orbitals. In Fig.~\ref{fig:Ti_projections} we have analyzed the projection of
        the Fermi surface states onto the Ti orbitals. The result is shown as a color scale on top of the Fermi surface, where the left end of the color scale
        (blue) corresponds to 10\% Ti contribution and the right end (red) to 90\%. Here one can see that those areas in Fig.~\ref{fig:Hall_contributions}
        that contribute most to the Hall coefficient have light blue color in Fig.~\ref{fig:Ti_projections}, which means that the Hall coefficient is
        predominantly coming from the Ni $d$-orbitals in both phases.

        In order to investigate the sensitivity of the Hall coefficient with respect to doping, we present in Fig.~\ref{fig:RH_vs_doping} the variation
        of $R_H$ with electron doping over a broad doping range. While there is some variation in the B19' phase, the variation in the B2 phase is
        relatively weak and $R_H$ remains negative throughout. We note that the doping level of our Ni$_{50.6}$Ti$_{49.4}$ sample would correspond to just 0.072 electrons/unit cell in
        the B2 phase. Thus, the doping effect cannot account for a sign change of $R_H$.

        In total we come to the conclusion that the Hall coefficient is not well represented by Boltzmann transport theory within
        the constant relaxation time approximation. In the B19' phase, while the temperature dependence of the resistivity can be reasonably reproduced,
        the temperature dependence of the Hall coefficient is not covered well. In the B2 phase the sign of the Hall coefficient is opposite
        to what is observed experimentally. The resistivity in the B2 phase is not well represented unless either extreme values for elastic scattering
        or a strong electron-phonon coupling strength would be assumed.

        \section{Charge density wave calculations}
        \label{sec:CDW}

        As pointed out in the introduction, there have been speculations about the formation of a CDW in NiTi in the past.
        Here, we will investigate the influence of different types of CDWs on the Hall coefficient to see whether they
        can lead to a better representation of the experimental results.

        \begin{figure}
		\includegraphics[width=1.0\linewidth]{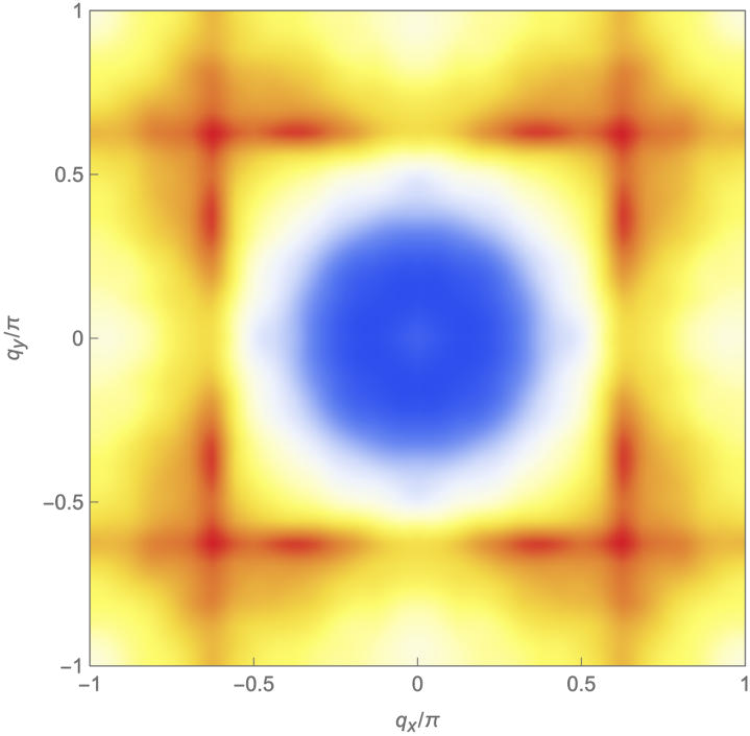}
		\caption{Cut of the Lindhard function in the $q_z=0$ plane. }
		\label{fig:Lindhard}
        \end{figure}

        Inelastic neutron scattering observed a softening of a transversal acoustic phonon mode at a wave vector
        of $Q=\frac{2}{3} (\pi,\pi,0)$ \cite{Tietze1984}. This wave vector is at odds with the unstable phonon modes
        seen at the wave vector $M=(\pi,\pi,0)$ in DFT calculations of the phonon modes \cite{Katsnelson2010,WuLawson2022}.
        It is also inconsistent with the doubling of the unit cell that occurs in the transition from the B2 to the B19' phase.
        Based on the wave vector $Q=\frac{2}{3} (\pi,\pi,0)$ a triplication of the unit cell might be expected. It has been suggested
        that this wave vector could be an indication of the formation of a CDW with a wave length of 3 lattice constants and might be stabilized
        by a strong electron-phonon interaction and a nesting feature of the Fermi surface \cite{ZhaoHarmon1993}.

        In order to check the nesting feature we have calculated the real part of the charge susceptibility (Lindhard function) using the
        {\it FermiSurfer} tool \cite{FermiSurfer}. In Fig.~\ref{fig:Lindhard} it is shown within the $q_z=0$ plane
        as a function of $q_x$ and $q_y$ for zero frequency. One can see square-like ridges, which occur at the
        wavevectors $q_x=\pm 0.625 (\pi,0,0)$ and $q_y=\pm 0.625 (0,\pi,0)$. A maximum is found at the
        wave vector $Q=0.625 (\pi,\pi,0)$. These features can be related to the nesting feature of the Fermi surface,
        which results from the cube-like structure in the B2 phase. In Fig.~\ref{fig:cutFS} we present a cut
        of the Fermi surface of the B2 phase in the $k_z=0$ plane. The black arrows show the
        vectors $Q_1=\frac{4}{3} (\pi,0,0)$ and $Q_2=\frac{4}{3} (0,\pi,0)$ that connect opposing sides of the cube. 
	Due to symmetry these vectors are equivalent to $\frac{2}{3} (\pi,0,0)$ and $\frac{2}{3} (0,\pi,0)$. As usual for real materials the nesting is not
        perfect. Correspondingly, the charge susceptibility does not diverge, but only shows a maximum at the
        nesting wave vectors.

        It has been pointed out by Johannes and Mazin \cite{JohannesMazin2008} that a nesting feature
        in the band structure does not necessarily mean that a CDW must occur at the same wavevector.
        There are examples of materials in which the nesting occurs at a very different wave vector
        than the CDW. The reason for such discrepancies is the possibility of strongly anisotropic
        electron-phonon interaction, which can stabilize CDWs at different wave vectors \cite{Weber2011,Zhu2015,Flicker2015,Gruenebohm2023}.
        In general a calculation of the momentum dependence of the electron-phonon interaction
        is necessary to get a complete picture of the CDW instability. As pointed out above,
        such calculations are complicated due to the entropic effects that need to be taken into account
        in the present case.

        Here, we will take a different, mean-field based approach to study the influence of
        different types of CDWs. Motivated by the above observations we will restrict ourselves
        to three specific commensurate types of CDWs that might occur in this material: as model A we consider
        a uniaxial CDW with a wave vector of $Q_A=\frac{2}{3} (\pi,\pi,0)$, model B is a biaxial
        CDW based on the two wave vectors $Q_{B1}=\frac{2}{3} (\pi,\pi,0)$ and $Q_{B2}=\frac{2}{3} (\pi,-\pi,0)$.
        The general treatment of these two kinds of CDWs and their influence on the Hall coefficient has been studied by Seo and Tewari as a
        model for cuprates \cite{SeoTewari2014}. Model A corresponds to a striped CDW, while model B describes a checkerboard
        type charge pattern. For comparison, as model C we also study a CDW with a wave vector of $Q_C=(\pi,\pi,0)$.
        The mean-field theory for this case was worked out by K.~Machida et al \cite{Machida1987,Machida1988,Ichimura1990,Dahm1997}
        for simple nearest-neighbor tight-binding models. Here we will generalize this to our realistic
        tight binding model of the B2 and the B19' phase of NiTi \cite{Dick2025}.

        \begin{figure}
		\includegraphics[width=1.0\linewidth]{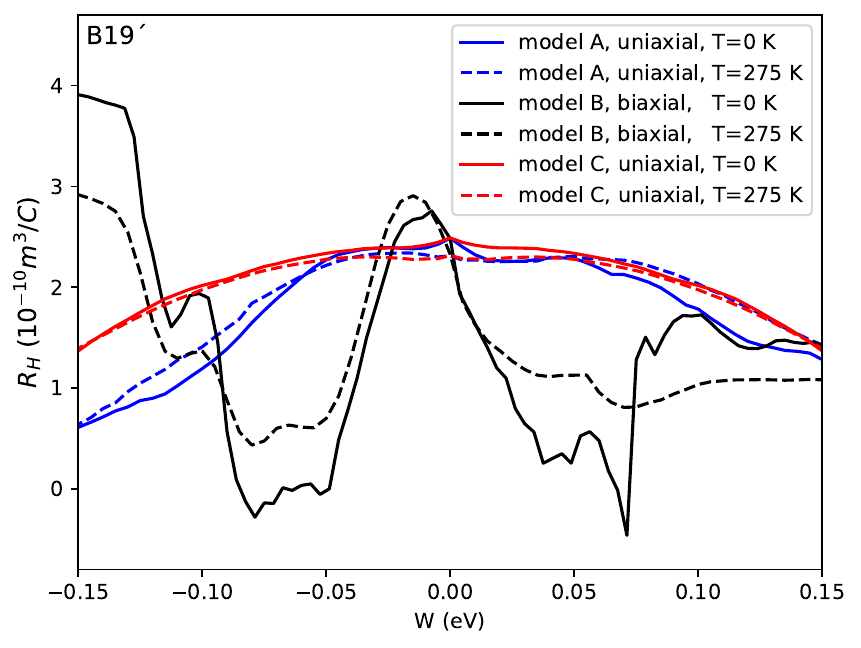}
		\includegraphics[width=1.0\linewidth]{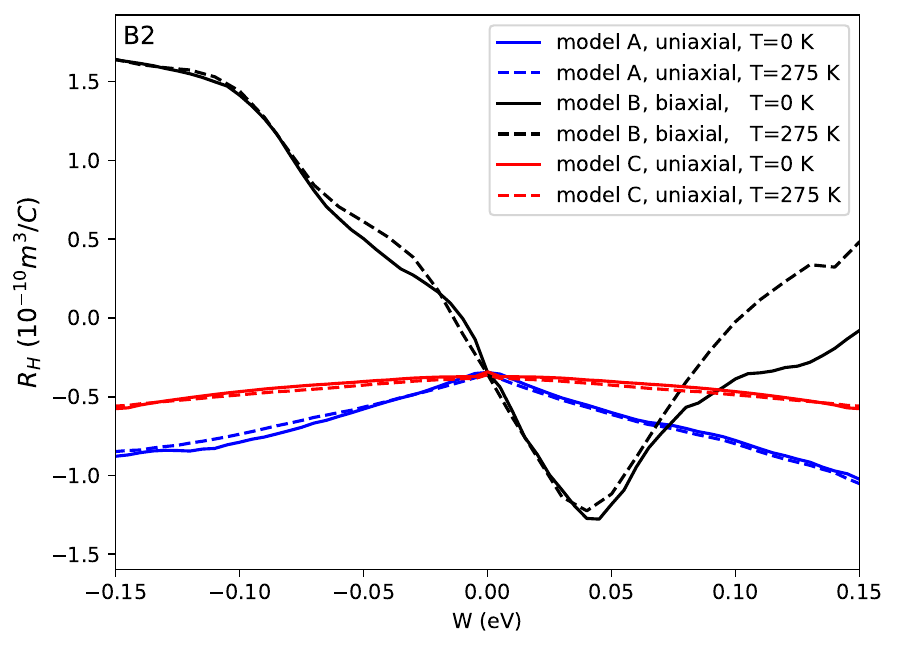}
		\caption{The Hall coefficient as a function of the CDW order parameter $W$ for the three models A (blue), B (black), and C (red).
                  The solid lines are for temperature $T=0$, while the dashed lines are for $T=275$~K. Panel (a) shows the results
                for the B19' phase and panel (b) for the B2 phase.}
		\label{fig:RH_W}
        \end{figure}

        Within mean-field theory, the Hamiltonian in the CDW phase is given by \cite{SeoTewari2014,Ichimura1990}
        \begin{equation}
          H_{MF} = \sum_{\mathbf{k}, n, \sigma} (\epsilon_{\mathbf{k} n} - \mu) c_{\mathbf{k} n \sigma}^{\dagger}c_{\mathbf{k} n \sigma} -
             \sum_{\mathbf{k}, \mathbf{Q}_i, n, \sigma} W_{\mathbf{Q}_i} c_{\mathbf{k}+\mathbf{Q}_i n \sigma}^{\dagger }c_{\mathbf{k} n \sigma}
          \label{eq:H_MF}
        \end{equation}
        with $W_{\mathbf{Q}_i}^*=W_{-\mathbf{Q}_i}$ such that $H_{MF}$ is hermitian. Here, $W_{\mathbf{Q}_i}$ is the CDW order parameter.
        The sum over $\mathbf{Q}_i$ runs over all nonzero
        and inequivalent linear combinations of the CDW wave vectors, i.e. for model A it runs over the two values $\pm Q_A$, for model B
        over the eight values $\pm Q_{B1}$, $\pm Q_{B2}$, and $\pm Q_{B1} \pm Q_{B2}$, and for model C just over the single value $Q_C$.
        We restrict the sum over the bands $n$ only to those bands that cross the Fermi energy, because CDWs typically possess a much smaller
        energy scale ($\sim 0.1$~eV) than the electronic band energies and thus only affect the states in the vicinity of the Fermi energy.
        Therefore, in the B2 phase we only consider the two bands shown in Fig.~\ref{fig:fermisurfaces}a and in the B19' phase the four
        bands shown in Fig.~\ref{fig:fermisurfaces}b.
        The sum over $\mathbf{k}$ is to be understood to run over the full first Brillouin zone of the original lattice without CDW.
        The presence of a commensurate CDW leads to a spontaneous breaking of the translational symmetry, which creates a larger
        unit cell. In model A this supercell is three times larger than the original unit cell, in model B nine times larger, and
        in model C twice as large. Correspondingly, the reduced Brillouin zone (RBZ) in model A/B/C is a factor of 3/9/2 smaller
        than the original Brillouin zone. In this situation it is useful to define a multi-component vector $\psi_{n,\sigma}(\mathbf{k})$
        with $\mathbf{k} \in$~RBZ from the operators $c_{\mathbf{k}+\mathbf{Q}_i n \sigma}$. This vector then possesses 3/9/2 components in model A/B/C.
        The mean-field Hamiltonian Eq.~(\ref{eq:H_MF}) can then be written in matrix form \cite{SeoTewari2014}
        \begin{equation}
          H_{MF} = \sum_{\mathbf{k} \in \mathrm{RBZ}, n, \sigma} \psi_{n,\sigma}(\mathbf{k})^\dagger H_n(\mathbf{k}) \psi_{n,\sigma}(\mathbf{k})
          \label{eq:H_MFmatrix}
        \end{equation}
        where $H_n(\mathbf{k})$ is a $3\times3$/$9\times9$/$2\times2$ matrix in model A/B/C. The diagonalization of this matrix yields the
        CDW subbands $E_{\mathbf{k} n}^{i}$ within the RBZ and the corresponding eigenstates. Having the CDW subbands, the Hall coefficient can be calculated
        again from Eq.~(\ref{eq:rank2_conductivity}), (\ref{eq:rank3_conductivity}), and (\ref{eq:rank3_Hall_tensor}), except that the calculation
        needs to be done for the CDW subbands $E_{\mathbf{k} n}^{i}$ instead
        of the original bandstructure $\epsilon_{\mathbf{k}n}$ \cite{SeoTewari2014}. In the following we will restrict ourselves to
        those CDW solutions with $W_{\mathbf{Q}_i}=W=$~const, assuming that all CDW components possess the same amplitude.

        Fig.~\ref{fig:RH_W} shows an overview on how the Hall coefficient depends on the CDW order parameter $W$ for the three models.
        Results are shown for temperature $T=0$ and for $T=275$~K near the transition. Fig.~\ref{fig:RH_W}a shows the behavior in
        the B19' phase and Fig.~\ref{fig:RH_W}b in the B2 phase. For model C one notes that the red curves are symmetric,
        i.e. $R_H(W)=R_H(-W)$, which is due to the higher symmetry of the wave vector $Q_C=(\pi,\pi,0)$. This symmetry is broken
        for the other two models. The two uniaxial models A and C lead to a decrease of $R_H$ as a function of increasing
        magnitude of the order parameter $|W|$. For the B19' phase this means that $R_H$ has its maximum value of $2.5 \cdot 10^{-10}$~m$^3$/C at $W=0$
        in model A and C. Thus, these models are unable to explain the large value of about $3 \cdot 10^{-10}$~m$^3$/C that is observed
        in the experimental data at low temperatures. In contrast, the biaxial model B allows values above $3 \cdot 10^{-10}$~m$^3$/C
        for negative $W$. Similarly in the B2 phase models A and C are unable to reach positive values of $R_H$. As it turns out to be
        impossible to understand the experimental result within model A or C, we will focus from now on on model B.
        
        \begin{figure}
		\includegraphics[width=1.0\linewidth]{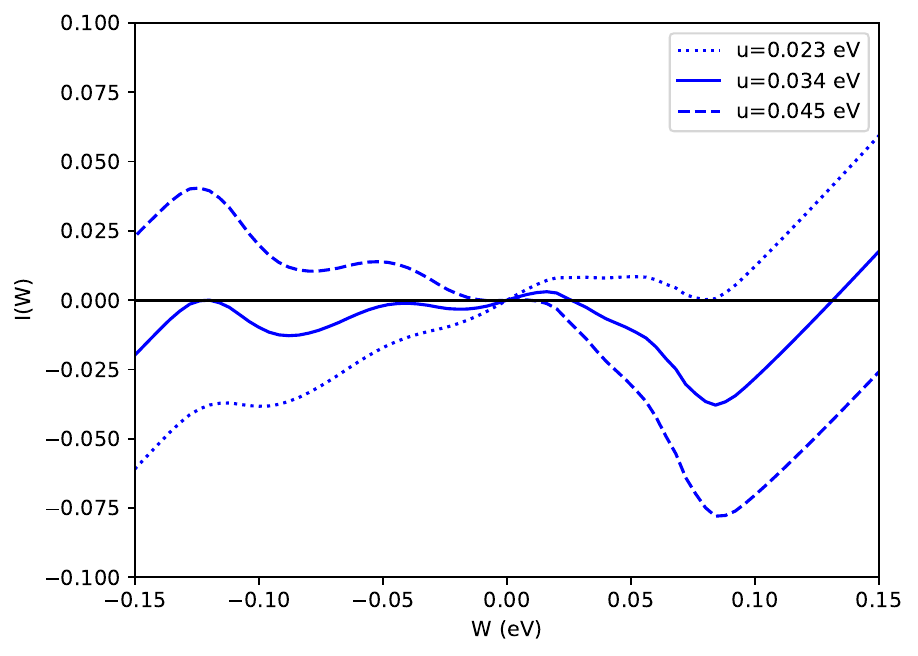}
		\caption{The function $I(W)$ (Eq.~(\ref{eq:I_W})) for model B in the B19' phase at $T=275$~K for three different values of the coupling constant $u$.}
		\label{fig:I_gap}
        \end{figure}

        The CDW order parameter $W$ has to be obtained from a self-consistency equation, which reads
        \begin{equation}
          W = u \sum_{\mathbf{k}, \mathbf{Q}_i, n, \sigma} \left\langle c_{\mathbf{k}+\mathbf{Q}_i n \sigma}^{\dagger }c_{\mathbf{k} n \sigma} \right\rangle_T
          \label{eq:SC_equation}
        \end{equation}
        where $\left\langle \cdots \right\rangle_T$ denotes the thermal expectation value at temperature $T$, which itself depends on $W$
        via the CDW subbands. Here, $u$ is the coupling constant for the CDW. In practice, we solve this self-consistency problem for $W$
        by numerically searching for zeroes of the function
        \begin{equation}
          I(W)= W - u \sum_{\mathbf{k}, \mathbf{Q}_i, n, \sigma} \left\langle c_{\mathbf{k}+\mathbf{Q}_i n \sigma}^{\dagger }c_{\mathbf{k} n \sigma} \right\rangle_T
          \label{eq:I_W}
        \end{equation}
        We note, that for each value of $W$ the chemical potential $\mu$ needs to be readjusted in such a way that the total number of
        electrons in the system is kept fixed \cite{SeoTewari2014}. Therefore, the search for the order parameter $W$ becomes a two parameter search for
        $W$ and $\mu$.

         \begin{figure}
		\includegraphics[width=1.0\linewidth]{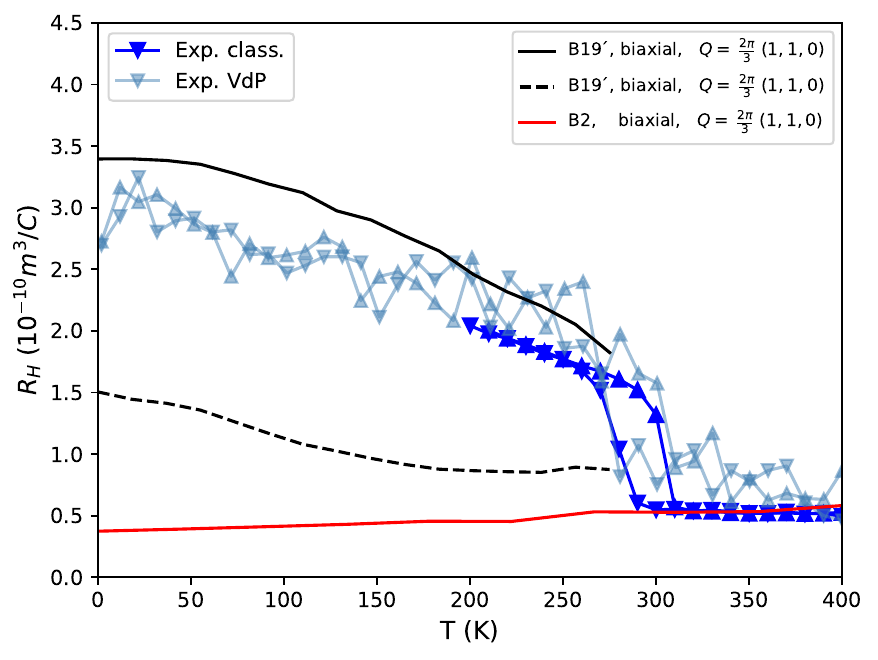}
		\caption{Temperature dependence of the Hall coefficient. Like in Fig.~\ref{fig:RHall_normal} the blue triangles show the experimental data on a Ni$_{50.6}$Ti$_{49.4}$ sample.
                  The black solid line shows the result for model B (biaxial CDW) and the solution with negative $W$. The black dashed line is
                  the result for model B's solution with positive $W$. The red line shows the result for model B in the B2 phase.}
		\label{fig:RHall_CDW}
        \end{figure}

       The function $I(W)$ is shown in Fig.~\ref{fig:I_gap} for model B at $T=275$~K for three different values of the coupling constant $u$. One can see that depending on
        the coupling strength $u$, $I(W)=0$ may possess several solutions. We point out that those solutions with $\frac{dI}{dW}<0$ are unstable
        as they correspond to local maxima of the free energy. There always exists the solution $W=0$ corresponding to the absence of a CDW.
        At $u=0.023$~eV a new stable solution at positive $W=0.082$~eV becomes possible. At $u=0.034$~eV another stable solution at negative $W=-0.121$~eV appears.
        To study these stable CDW solutions, we choose $u=0.023$~eV and $u=0.034$~eV, obtain the temperature
        dependence of $W$ from Eq.~(\ref{eq:I_W}), and calculate the CDW subbands and the Hall coefficient $R_H$ as a function of temperature.

        In Fig.~\ref{fig:RHall_CDW} we show the temperature dependence of the Hall coefficient $R_H$ that results in the B19' (black) and B2 (red) phases for model B.
        The result for the solution with negative $W<0$ is shown as the solid black line and the other solution with positive $W>0$ as the dashed solid line. As already
        anticipated from Fig.~\ref{fig:RH_W}a the positive solution results in a Hall coefficient that is too small compared with the data. The negative
        solution, however, is in reasonable agreement with the data.
        In contrast to the solution without CDW (Fig.~\ref{fig:RHall_normal}) the Hall coefficient
        now increases with decreasing temperature in the B19' phase. This results from the temperature dependence of the order parameter $W$.

         \begin{figure}
		\includegraphics[width=1.0\linewidth]{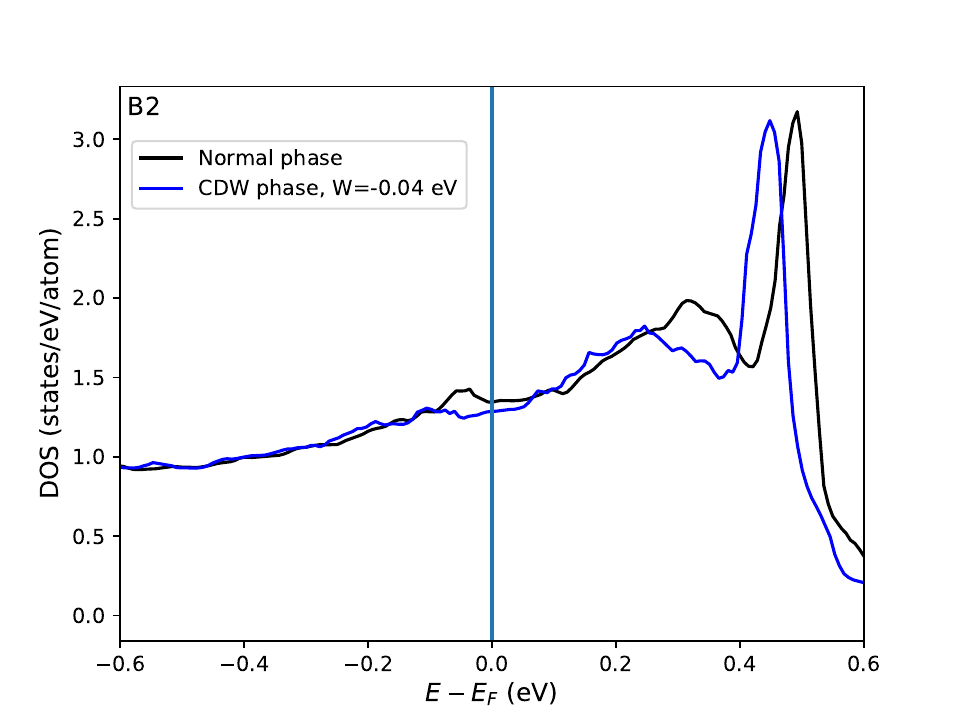}
		\includegraphics[width=1.0\linewidth]{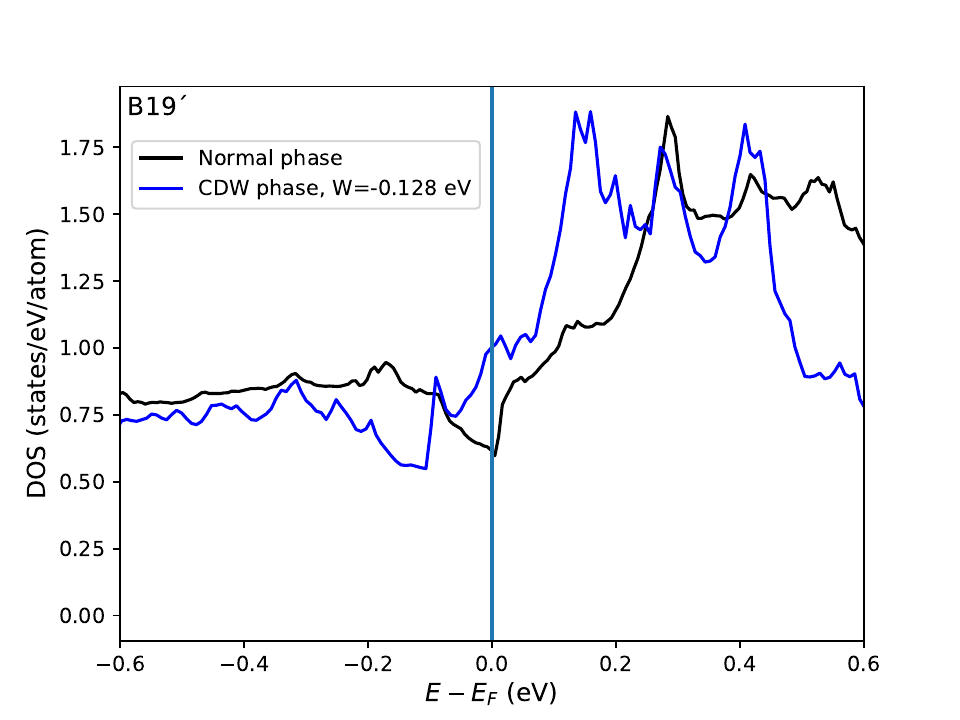}
		\caption{Density of states at $T=0$ with (blue line) and without CDW (black line). Panel (a) shows the B2 phase and (b) the B19' phase.}
		\label{fig:dos_CDW}
        \end{figure}
        
       In the B2 phase we have fixed the coupling constant $u$ such that at 350~K the Hall coefficient in the data is reproduced. In this way we found
        $u=0.041$~eV yielding $W=-0.038$~eV.
        We have solved the self-consistency equation (\ref{eq:SC_equation}) to determine the temperature dependence of $W$, which turned out to be weak in this case. The resulting temperature dependence of $R_H$
        is shown as the red curve. In contrast to the solution without CDW a positive sign of $R_H$ in agreement with the data is now obtained.
        The picture that is suggested by this analysis would be that in the B2 phase a biaxial CDW with a small amplitude or possibly a fluctuating precursor CDW
        is already present. When the transition to the B19' phase sets in, this CDW is boosted to a larger amplitude.

        Fig.~\ref{fig:dos_CDW} shows the density of states at zero temperature in the vicinity of the Fermi surface with and without CDW for the two phases.
        In the B2 phase there is only a small reduction of the density of states at the Fermi energy due to the small size of the order parameter $W$. In
        the B19' phase, however, we find a visible increase of the density of states at the Fermi energy. Note that the peak in the density of states
        present at $\sim0.3$~eV in the absence of the CDW is splitted into three peaks in the CDW phase as a result of the formation of the CDW subbands.
        As a side effect of that splitting the density of states at the Fermi energy is lifted.
        
        \begin{figure}[t]
		\includegraphics[width=1.0\linewidth]{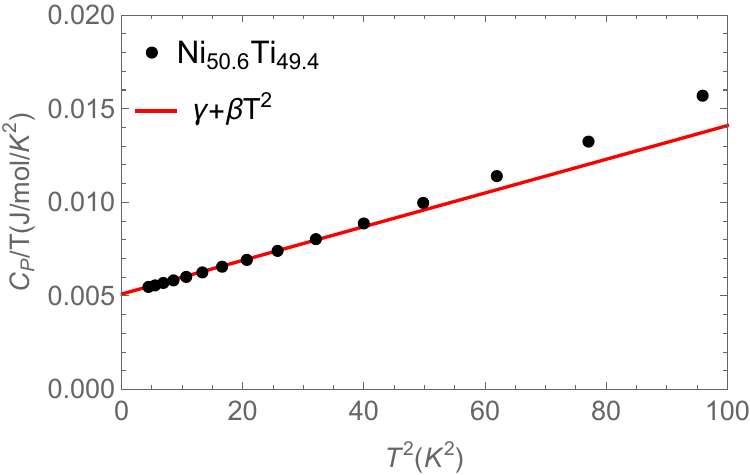}
		\caption{Low temperature specific heat data of Ni$_{50.6}$Ti$_{49.4}$ from Ref.~\onlinecite{KunzmannDiss} (black dots) presented as $C_P/T$ versus $T^2$.
                  The red line shows the
                  relationship $\gamma + \beta T^2$ with $\gamma=5.1\cdot 10^{-3}$~J/mol/K$^2$ and $\beta=9.0 \cdot 10^{-5}$~J/mol/K$^4$.}
		\label{fig:specific_heat_low_temp}
        \end{figure}

        The density of states at the Fermi level is related to the electronic contribution to the specific heat. At low temperatures the specific heat
        follows the law \cite{Czycholl2023}
        \begin{equation}
          C_V = \gamma T + \beta T^3
          \label{eq:specific_heat_low_temp}
        \end{equation}
        The contribution third order in $T$ is due to the acoustic phonons of the material, while the term linear in $T$ is due to the electrons.
        In Fig.~\ref{fig:specific_heat_low_temp} we demonstrate that the data from Ref.~\onlinecite{KunzmannDiss} (black dots) follow this relationship
        with the values $\gamma_{exp}=5.1\cdot 10^{-3}$~J/mol/K$^2$ and $\beta_{exp}=9.0 \cdot 10^{-5}$~J/mol/K$^4$.
        
        The coefficient $\gamma$ is proportional to the density of states at the Fermi level $D(E_F)$ \cite{Czycholl2023}:
        \begin{equation}
          \gamma = \frac{\pi^2}{3} k_B^2 D(E_F)
          \label{eq:specific_heat_coefficient}
        \end{equation}
        We can thus use our theoretical values of the density of states in the B19' phase in Fig.~\ref{fig:dos_CDW}b to obtain theoretical values for
        the specific heat coefficient $\gamma$. In the CDW phase we have $D_{CDW}(E_F)=1.006$~states/eV/atom while without CDW we have
        $D_{0}(E_F)=0.618$~states/eV/atom. From this we find $\gamma_0=2.9\cdot 10^{-3}$~J/mol/K$^2$ and $\gamma_{CDW}=4.7\cdot 10^{-3}$~J/mol/K$^2$, which
        shows that the density of states in the CDW state is consistent with the specific heat data, while the DFT result without CDW is somewhat too small.


        \section{Conclusions}

        We have shown that the experimental data of the Hall coefficient in NiTi cannot be well understood within Boltzmann transport theory based on DFT bandstructure.
        The temperature dependence of the resistivity can be understood within the martensite phase, however the behaviour in the austenite phase suggests a significant
        increase of the electron-phonon coupling strength. We have studied the influence of three specific kinds of CDWs on the Hall coefficient using a mean-field CDW theory.
        We found that uniaxial CDWs cannot represent the Hall data well, however, a biaxial CDW solution was found to be in reasonable agreement with the data. The analysis suggests
        that a small amplitude precursor CDW may already be present in the austenite phase and gets boosted when the martensite phase is entered.
        In contrast to a naive expectation, the density of states at the Fermi energy in the martensite phase increases in the presence of the biaxial CDW. We showed that this increased
        density of states is actually quantitatively consistent with low temperature specific heat data.

        \section{Acknowledgements}
        We thank Anna B\"ohmer, Ralf Drautz, Ilya Eremin, Jan Frenzel, Rolf Heid, Martin Mittendorff, Malte R\"osner, Kai Rossnagel, and Frank Weber for valuable discussions.
        The Physical Property Measurement System was funded by a large-scale equipment grant from the Deutsche Forschungsgemeinschaft (DFG, German Research Foundation) 
	under Article 91 b under grant agreement number INST 215/619-1 FUGG, gratefully acknowledged by G.S. 
	We also thank the HPC.NRW team for their support, as some of the DFT calculations have been done on the Bielefeld GPU Cluster.
        
        
	\bibliographystyle{apsrev4-2}
	\bibliography{bibliography}

        
\end{document}


\title{Influence of Charge Density Waves on the Hall coefficient in NiTi \\ Supplementary information}
	\author{Adrian Braun}
	\affiliation{
		Physics Department, Bielefeld University, Postfach 100131, 33501 Bielefeld, Germany
	}
	\author{Henrik Dick}
	\affiliation{
		Physics Department, Bielefeld University, Postfach 100131, 33501 Bielefeld, Germany
	}
	\author{Timon Sieweke}
	\affiliation{
		Institute for Energy and Materials Processes – Applied Quantum Materials, University Duisburg-Essen, 47057 Duisburg, Germany
	}
	\author{Alexander Kunzmann}
	\affiliation{
		Institute for Energy and Materials Processes – Applied Quantum Materials, University Duisburg-Essen, 47057 Duisburg, Germany
	}
	\author{Klara L\"unser}
	\affiliation{
		Institute for Energy and Materials Processes – Applied Quantum Materials, University Duisburg-Essen, 47057 Duisburg, Germany
	}
 	\author{Gabi Schierning}
	\affiliation{
		Institute for Energy and Materials Processes – Applied Quantum Materials, University Duisburg-Essen, 47057 Duisburg, Germany
	}
        \author{Thomas Dahm}
	\affiliation{
		Physics Department, Bielefeld University, Postfach 100131, 33501 Bielefeld, Germany
	}
	
	\date{\today}
	
	\maketitle
	\section{Extraction of the Debye temperature from specific heat data}
	\label{sec1}

        \begin{figure}
		\includegraphics[width=1.0\linewidth]{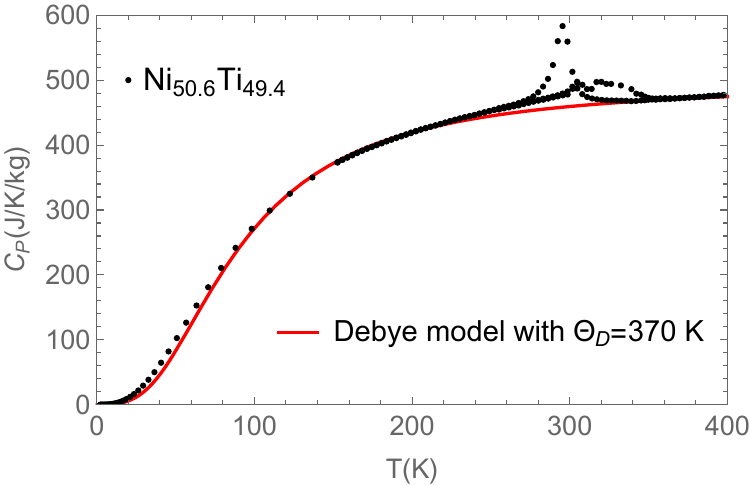}
		\caption{Specific heat data from Ref.~\onlinecite{KunzmannDiss} (black dots) together with a fit to the Debye model with $\Theta_D=370$~K (red line).}
		\label{fig:specific_heat}
        \end{figure}

        To obtain the Debye temperature $\Theta_D$, we fit the specific heat data from Ref.~\onlinecite{KunzmannDiss} to the Debye model
         \begin{equation}
          C_V(T) = 3 C_{\infty} \left( \frac{T}{\Theta_D} \right)^3 \int_{0}^{\Theta_D/T} dx \; \frac{x^4 e^x}{\left( e^x - 1 \right)^2}
          \label{eq:debye_model}
        \end{equation}
         where $C_{\infty}$ is the limiting value at high temperatures. The fit is shown in Fig.~\ref{fig:specific_heat} and was obtained for $\Theta_D=370$~K
         and $C_{\infty}=495$~J/K/kg.

         Alternatively one can also obtain a value of the Debye temperature $\Theta_{D,low T}$ from the low temperature coefficient $\beta=9.0 \cdot 10^{-5}$~J/mol/K$^4$
         found in Fig.~16 of the main manuscript using the relationship \cite{Czycholl2023}
         \begin{equation}
          \Theta_{D,low T} = \sqrt[3]{\frac{12 \pi^4 r R}{5 \beta}}
          \label{eq:debye_low_temp}
         \end{equation}
         where $R$ is the molar gas constant and the factor $r=2$ the number of atoms per unit cell in NiTi. This yields $\Theta_{D,low T}=351$~K.
         We point out that this value describes the acoustic phonon modes, while the above value of 370~K is an
         effective value of the complete phonon spectrum including acoustic and optical phonon modes.

	\section{Electronic entropy}
	\label{sec2}

        \begin{figure}
		\includegraphics[width=1.0\linewidth]{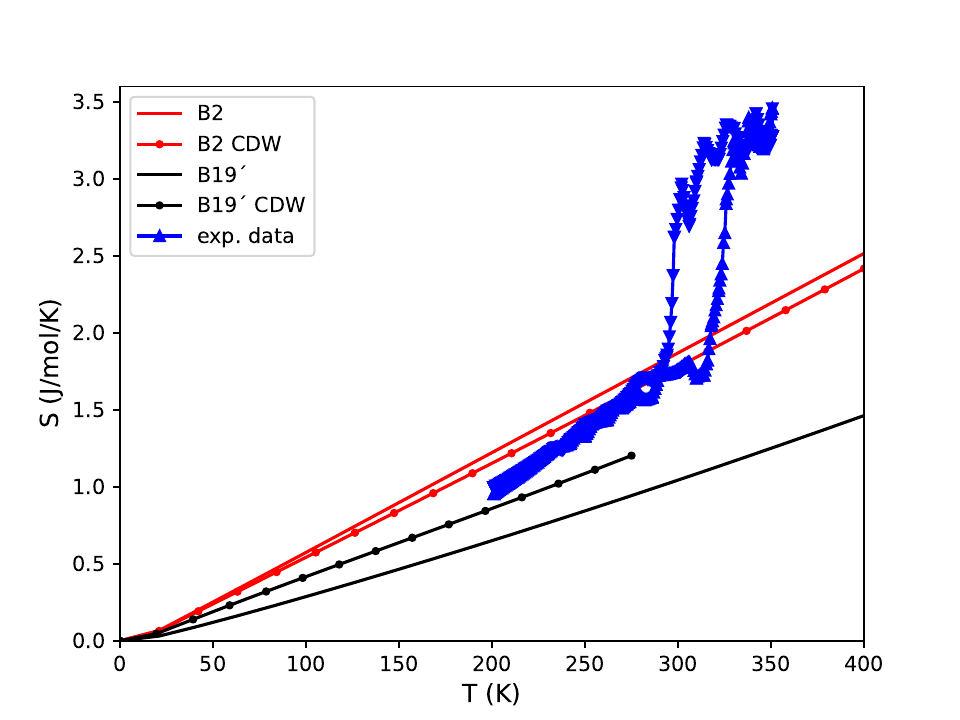}
		\caption{Electronic entropy as a function of temperature in the B2 (red) and B19' (black) phases with (circles) and without CDW (no symbols). The blue triangles show
               experimental data from Ref.~\onlinecite{Kunzmann2022} for Ni$_{50.6}$Ti$_{49.4}$.}
		\label{fig:electronic_entropy}
        \end{figure}

        In Ref.~\onlinecite{Kunzmann2022,KunzmannDiss} it was found that the change in the electronic entropy at the martensitic transition is surprisingly large, suggesting a large
        contribution of the electronic system to the phase transition. In order to address this observation, we also calculated the
        electronic part of the entropy for our model.

\begin{figure}
		\includegraphics[width=1.0\linewidth]{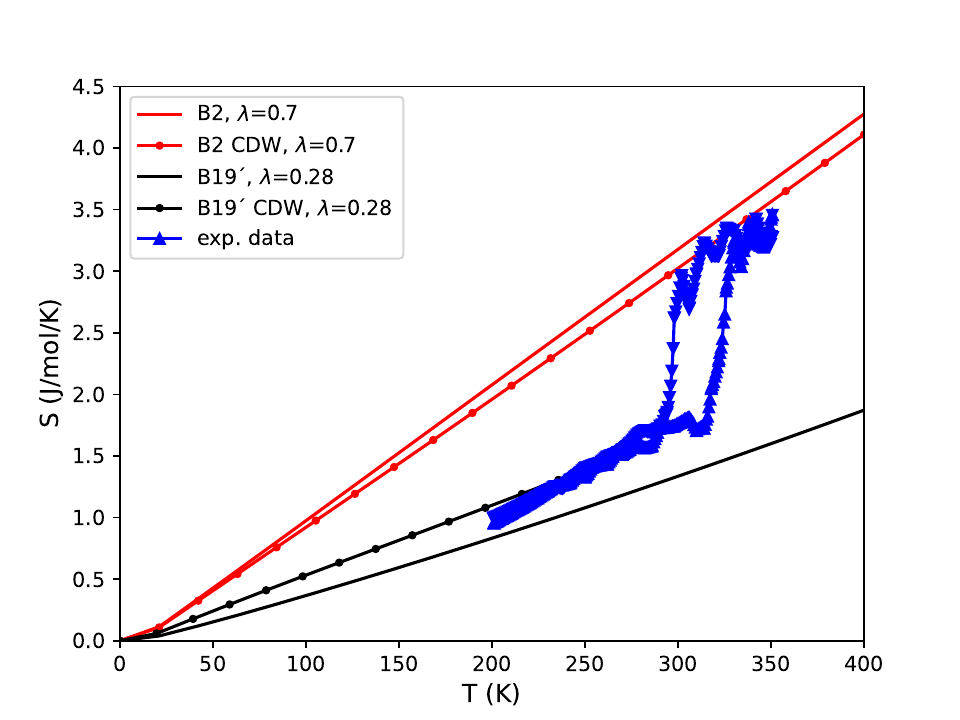}
		\caption{Same as Fig.\ref{fig:electronic_entropy} but when the mass renormalization
                  effect due to electron-phonon coupling is taken into account.}
		\label{fig:electronic_entropy_scaled}
        \end{figure}

        We first calculate the total energy $U(T)$ of the electronic system using the expression
        \begin{equation}
          U(T) = \int_{-\infty}^{\infty} d\epsilon \; D(\epsilon) \epsilon f(\epsilon) \; .
          \label{eq:electronic_energy}
        \end{equation}
	Here, $D(\epsilon)$ is the density of states relative to the Fermi level and $f(\epsilon)$ the Fermi function.
	Note, that both the Fermi function as well as the density of states depend on temperature, because the CDW order parameter $W(T)$
	is temperature dependent. From Eq.~(\ref{eq:electronic_energy}) the specific heat $C_V$ and the entropy $S$ are obtained from the
	expressions
        \begin{equation}
          C_V = \frac{dU}{dT} \qquad \mbox{and} \qquad S = \int_0^T \frac{dT'}{T'} C_V(T')
          \label{eq:specific_heat_entropy}
        \end{equation}
	The result of this calculation in both phases B2 and B19' both with and without CDW is shown in Fig.~\ref{fig:electronic_entropy} 
	along with experimental data from Ref.~\onlinecite{KunzmannDiss}. One can see that the influence of the CDW on the electronic entropy
	is only moderate due to its moderate effect on the density of states discussed in sec~IV of the main manuscript. Generally, the comparison with the experimental
 	data is not very convincing. We point out, however, that the influence of electron-phonon interaction is not included in Fig.~\ref{fig:electronic_entropy}.
	Electron-phonon interaction leads to an enhancement of the effective band mass of the electrons at the Fermi energy of the form 
	$m^*=m (1+\lambda)$ \cite{Czycholl2023}, where $\lambda$ is the so-called mass enhancement factor \cite{Allen1983,Carbotte1990}, which
	is a dimensionless measure of the electron-phonon coupling strength. This leads to an increase of the electronic entropy
	by the factor $1+\lambda$. $\lambda$ is related, but in general not equal to the
	transport coupling strength $\lambda_{tr}$, which we determined in Sec.~III of the main mansucript. If we assume $\lambda \approx \lambda_{tr}$, we
	can estimate the influence of electron-phonon coupling onto the electronic entropy. In Fig.~\ref{fig:electronic_entropy_scaled}
	we assumed $\lambda=0.28$ in the B19' phase and $\lambda=0.7$ in the B2 phase, the values we found in Sec.~III, main manuscript.
	Interestingly, the agreement with the experimental data becomes very good, if these coupling strengths are assumed.
	This suggests that the electron-phonon interaction strength is significantly larger in the B2 phase than in the B19'
	phase, in line with the analysis of the resistivity data in Sec.~III, main manuscript. Thus, we can understand a strong change in electronic entropy, which
	was discussed in Ref.~\onlinecite{Kunzmann2022}, as a consequence of a strong change of electron-phonon interaction strength and
	the related change in effective mass of the electrons. We wish to caution, however, that the electronic entropy in Ref.~\onlinecite{Kunzmann2022}
        was not measured directly, but obtained indirectly via the Seebeck coefficient, so this comparison should be taken with some care.

	\section{Numerical evaluation of transport coefficients}
	\label{sec3}

        Physical quantities like the Hall coefficient require an integration in $\mathbf{k}$-space within the Brillouin zone ($BZ$) of the examined lattice.
        In this work the tetrahedron method, as described in \cite{Tetrahedron1994}, was performed, where the integration was carried out over the whole $BZ$ with application of the correction terms
        discussed in Ref.~\onlinecite{Tetrahedron1994}.\\
The size of the reduced $BZ$ of the super lattice ($BZ_{CDW}$) depends on the examined CDW vector $\bm{Q}$ and is $N$-times smaller than the $BZ$ in the normal phase without CDW.\\

\begin{table}[h!]
\caption{CDW type and volume ratio N=$V_{BZ} / V_{BZ_{CDW}}$}
\centering
\begin{tabular}{|c||c||c|}

\hline
CDW type & CDW vector $\bm{Q}$  & N\\
\hline
uniaxial & $\bm{Q}=\pi(1,1,0)$               & 2 \\
uniaxial & $\bm{Q}=\frac{2}{3}\pi(1,1,0)$    & 3 \\
biaxial  & $\bm{Q}_1= \frac{2}{3}\pi(1,1,0)$ & 9\\
         & $\bm{Q}_2=\frac{2}{3}\pi(1,-1,0)$ & \\
\hline
\end{tabular}
\label{tab1}
\end{table}

In order to calculate the energy eigenvalues in the charge density wave (CDW) phase, the bands are folded into $BZ_{CDW}$.
Formally this was done by evaluating the energy eigenvalues $\epsilon_n(\mathbf{k})$ of the normal phase in $BZ_{CDW}$ and all regions within $BZ$ that result from a shift of $BZ_{CDW}$  by the corresponding CDW vectors $\bm{Q}$. In the case of $\bm{Q}=\frac{2}{3}\pi(1,1,0)$ for example,
\begin{align}
H = \begin{pmatrix} \epsilon_{\bm{k}}  & -W & -W\\
-W  & \epsilon_{\bm{k}+\bm{Q}} & -W\\
-W  & -W & \epsilon_{\bm{k}-\bm{Q}}\\
\end{pmatrix} - \mu \mathbbold{1}
\end{align}
has to be diagonalized for each $\mathbf{k} \in BZ_{CDW}$ to obtain the CDW bands $E_{n_{CDW}}(\mathbf{k}, W)$, where W denotes the CDW order parameter. In the same way, the first and second derivative of the bands in the normal phase are calculated, from which $\nabla_{\mathbf{k}} H$ and $\Delta_{\mathbf{k}} H$ in the CDW phase can be obtained. Note that the off diagonal elements of $\nabla_{\mathbf{k}} H$ and $\Delta_{\mathbf{k}} H$  are zero, since the order parameter W is assumed $\bm{k}$-independent in this case. The group velocity and inverse mass tensor in the CDW case are then determined by first and second order perturbation theory using H, $\nabla_{\mathbf{k}} H$ and $\Delta_{\mathbf{k}} H$. Under the assumption of no degeneracy, the first derivatives of the bands $E_d$ are calculated by:\\
\begin{align}
\frac{\partial E_d}{\partial k_i} = \frac{\partial E_d}{\partial H_{nm}} \frac{\partial H_{nm}}{\partial k_i}
\end{align}
with
\begin{align}
\frac{\partial E_d}{\partial H_{nm}}=U_{md} U_{dn}^{-1} \quad \text{ and } \quad  H_{nm} = U_{nd} E_d U_{dm}^{-1}
\end{align}
where $U_{nm}$ is the unitary matrix which diagonalizes $H$.
The matrix elements of $\Delta_{\mathbf{k}} H$ are calculated via
\begin{align}
\frac{\partial^2 E_d}{\partial k_i \partial k_j} = \frac{\partial^2 E_d}{\partial H_{nm} \partial H_{kl}} \frac{\partial H_{nm}}{\partial k_i} \frac{\partial H_{kl}}{\partial k_j} + \frac{\partial E_d}{\partial H_{nm}} \frac{\partial^2 H_{nm}}{\partial{k_i} \partial{k_j}}
\end{align}
Using
\begin{align}
\frac{\partial U_{md}}{\partial H_{kl}} = \sum_{j \neq d} \frac{U_{mj} \overline{U_{kj}} U_{ld}}{E_d - E_j}
\end{align}
and 
\begin{align}
\frac{\partial^2 E_d}{\partial H_{nm} \partial H_{kl}} = 2 \frac{\partial U_{md}}{\partial H_{kl}} \overline{U_{nd}} = 2 \sum_{j \neq d} \frac{U_{mj} \overline{U_{kj}} U_{ld} \overline{U_{nd}}}{E_d - E_j}
\end{align}
the second derivatives can be re-expressed as:
\begin{align}
\frac{\partial^2 E_d}{\partial k_i \partial k_j} = 2 \sum_{j \neq d} \frac{U_{mj} \overline{U_{kj}} U_{ld} \overline{U_{nd}}}{E_d - E_j} \frac{\partial H_{nm}}{\partial k_i} \frac{\partial H_{kl}}{\partial k_j} + U_{nd} \overline{U_{md}} \frac{\partial H_{nm}}{\partial k_i \partial k_j}
\end{align}
For degenerate bands, additional steps are required.\\
The sum of an arbitrary function that is periodic in the reciprocal basis vectors in $\mathbf{k}$-space in the normal phase is equal to a sum within $BZ_{CDW}$ of all subbands resulting from the bandfolding into $BZ_{CDW}$.
\begin{align}
\sum_{\mathbf{k}, \sigma, n}  f(\mathbf{k}) = 2 \sum_{\mathbf{k}, n}  f(\mathbf{k}) = 2 \sum_{\mathbf{k}, n_{CDW}}^{'}  f(\mathbf{k})
\end{align}
for $W=0$. Here $\sum_{\mathbf{k}}$ stands for a summation over $BZ$ and $\sum_{\mathbf{k}}^{'}$ for one over $BZ_{CDW}$, where n denotes the band indices of those bands in normal phase in which a CDW order parameter is assumed to be present, while $n_{CDW}$ represents the band indices in the CDW phase. Since $BZ$ is N times larger than $BZ_{CDW}$, there are $n*N$ bands in the CDW phase.

\subsection{Blöchl weights}
The equation for the CDW order parameter is of the form
\begin{align}
\braket{X} &= \frac{1}{V_G} \sum_n \int_{V_G} d^3 k \: X_n(\mathbf{k}) \: f(\epsilon_n (\mathbf{k})) \label{gapequation}\\
&\overset{\mathrm{T \rightarrow 0}}{=} \frac{1}{V_G} \sum_n \int_{V_G} d^3 k \: X_n(\mathbf{k}) \: \left( 1 - \theta(\epsilon_n (\mathbf{k}) - \mu) \right)
\end{align}
In the numerical calculation, this can be rewritten as
\begin{align}
\braket{X} = \sum_{Tet} \sum_{j,n}  \: X_n(\mathbf{k}_{j}) \: \omega_{nj, Tet}
\end{align}
where $\sum_{Tet}$ means the summation over all tetrahedra and $\omega_{nj, Tet} = \tilde{\omega}_{nj, Tet} + \omega_{nj, Tet, corr}$ denote the integration weights of the tetrahedron corners. $\tilde{\omega}_{nj, Tet}$ resembles the weights without correction and $\omega_{nj, Tet, corr}$ the correction term \cite{Tetrahedron1994}, which reduces the numerical error by accounting for the band's curvature. Note, that the weights are independent of temperature, since the temperature is set to zero in this numerical formalism.
Similarly, the expressions for the Hall coefficient and the Drude weight are of the following form:
\begin{align}\label{Blöchl_Hall}
\braket{X} &= \frac{1}{V_G} \sum_n \int_{V_G} d^3 k \: X_n(\mathbf{k}) \: \left(-\frac{\partial}{\partial \epsilon} f(\epsilon_n (\mathbf{k})) \right)\\
&\overset{\mathrm{T \rightarrow 0}}{=}  \frac{1}{V_G} \sum_n \int_{V_G} d^3 k \: X_n(\mathbf{k}) \: \delta (\epsilon_n (\mathbf{k}) - \mu)
\end{align}
which is numerically evaluated as
\begin{align}
\braket{X} = \sum_{Tet} \sum_{j,n}  \: X_n(\mathbf{k}_{j}) \: d\omega_{nj, Tet}
\end{align}
Since the Fermi function is substituted by the integration weight $\omega_{nj}$, the weights for integrands like in \eqref{Blöchl_Hall} can be obtained as the derivative of the integration weights with respect to $\mu$:
\begin{align}
d \omega_{nj, Tet} = \frac{\partial}{\partial \mu} \omega_{nj, Tet}
\end{align}

\subsection{Temperature dependence}

In order to calculate the temperature dependence, it is useful to transform the integral in $\mathbf{k}$-space into one over an energy domain:
\begin{align}\label{formal_T_function}
\braket{X} &= \frac{1}{V_G} \sum_n \int_{V_G} d^3 k \: X_n(\mathbf{k}) \: \left(-\frac{\partial}{\partial \epsilon} f(\epsilon_n (\mathbf{k})) \right)\\
&= \int_{-\infty}^{\infty} d\epsilon \: \frac{1}{V_G} \sum_n \int_{V_G} d^3 k \: \delta(\epsilon_n (\mathbf{k}) - \epsilon) \: X_n(\mathbf{k}) \: \left(-\frac{\partial}{\partial \epsilon} f(\epsilon_n (\mathbf{k})) \right)\\
&= \int_{-\infty}^{\infty} d\epsilon \left(-\frac{\partial}{\partial \epsilon} f(\epsilon) \right) \frac{1}{V_G} \sum_n \int_{V_G} d^3 k \: \delta(\epsilon_n (\mathbf{k}) - \epsilon) \: X_n(\mathbf{k})
\end{align}
Temperature dependent quantities of the form
\begin{align}\label{K_integral}
\int_{-\infty}^{\infty} d\epsilon \: K(\epsilon) \: \left(- \frac{d}{d\epsilon} f(\epsilon) \right)
\end{align}
can be approximated as:
\begin{align}
\int_{\mu - 10 k_B T}^{\mu + 10 k_B T} d\epsilon \: K(\epsilon) \: \left(- \frac{d}{d\epsilon} f(\epsilon) \right) = \int_{\mu - 10 k_B T}^{\mu + 10 k_B T} d\epsilon \frac{K(\epsilon) e^{\frac{\epsilon - \mu}{k_B T}}}{k_B T \left( 1 + e^{\frac{\epsilon - \mu}{k_B T}} \right)^2} 
\end{align}
since the integrand only contributes within this range, due to the derivative of the Fermi function. $K(\epsilon)$ is piecewise linearized over a set of discrete $\epsilon$-values within $[\mu - 10 k_B T, \mu + 10 k_B T]$. The analytical integration of
\begin{align}
\sum_{\mu - 10 k_B T < \epsilon_l < \mu + 10 k_B T} \int_{\epsilon_l}^{\epsilon_{l+1}} d\epsilon \: \frac{\left( K(\epsilon_i) + \frac{(\epsilon - \epsilon_i)(K(\epsilon_{i+1}) - K(\epsilon_i))}{(\epsilon_{i+1} - \epsilon_i)} \right) e^{\frac{\epsilon - \mu}{k_B T}}}{k_B T \left( 1 + e^{\frac{\epsilon - \mu}{k_B T}} \right)^2}
\end{align}
is
\begin{align}
\sum_{\mu - 10 k_B T < \epsilon_l < \mu + 10 k_B T} & \frac{\frac{K(\epsilon_{i+1}) - K(\epsilon_i)}{(\epsilon_{i+1} - \epsilon_i)} \left( e^{\frac{\epsilon - \mu}{k_B T}} (\epsilon - \mu) + (\epsilon_i - \mu) \right) - K(\epsilon_i)}{1 + e^{\frac{\epsilon - \mu}{k_B T}}} \nonumber\\
& - k_B T \frac{K(\epsilon_{i+1}) - K(\epsilon_i)}{(\epsilon_{i+1} - \epsilon_i)} ln\left( 1 + e^{\frac{\epsilon - \mu}{k_B T}} \right) \bigg\rvert_{\epsilon_l}^{\epsilon_{l+1}}
\end{align}
For
\begin{align}
K(\epsilon)= \int_{-\infty}^{\epsilon} D(\epsilon^{'}) d\epsilon^{'}
\end{align}
the above integral (\ref{K_integral}) describes the particle number n and is used to calculate the chemical potential for any temperature, by means of a root finding algorithm. $\mu$ is varied such, that the particle number is equal to a fixed value for a given T and order parameter W. When considering a CDW phase, the number of bands within $BZ_{CDW}$ is N times larger than the initially incorporated number of bands of the normal phase. Therefore the energy surfaces resulting from the above integration over e intersect only with a smaller subset of all CDW energy bands. Thus those bands, which do not contribute to a certain energy surface, are neglected in the numerical calculation. The chemical potential was recalculated for all T and W in the scheme of the piecewise linearization, since it improved the result and accuracy. Although the equation for the CDW order parameter \eqref{gapequation} contains the Fermi function in the integrand, it can also be brought into the form of \eqref{formal_T_function} by partial integration in order to use the described procedure.

	\bibliographystyle{apsrev4-2}
	\bibliography{bibliography}